\documentstyle[aas2pp4]{article}

\begin{document}
\title{SBS~0335--052W -- AN EXTREMELY LOW METALLICITY DWARF 
GALAXY{\footnote[1]{Some of the observations reported here were obtained at 
the Multiple Mirror Telescope Observatory, a joint facility of the Smithsonian
Institution and the University of Arizona, and at the W. M. Keck
Observatory (WMKO), which is operated jointly by the California Institute of
Technology, the University of California, and NASA. WMKO was made possible
by the generous support of the W. M. Keck Foundation.}}}
\author{Valentin A. Lipovetsky\footnote[2]{Visiting astronomer, 
National Optical Astronomical Observatories.}$^,$\footnote[3]{Deceased 1996.}}
\affil{Special Astrophysical Observatory, Russian Academy of Sciences,
Nizhny Arkhyz, Karachai-Circessia 357147, Russia \\ Electronic mail:
val@sao.ru}
and
\author{Frederic H. Chaffee}
\affil{W. M. Keck Observatory, 65-1120 Mamalahoa Hwy., Kamuela,
HI 96743 \\ Electronic mail: fchaffee@keck.hawaii.edu}
\and
\author{Yuri I. Izotov{\footnotemark[2]}}
\affil{Main Astronomical Observatory, Ukrainian National Academy of Sciences,
Goloseevo, Kiev 252650, Ukraine \\ Electronic mail: izotov@mao.kiev.ua}
\and
\author{Craig B. Foltz}
\affil{Multiple Mirror Telescope Observatory, University of Arizona, 
Tucson, AZ 85721 \\ Electronic mail: cfoltz@as.arizona.edu}
\and
\author{Alexei Y. Kniazev{\footnotemark[2]}}
\affil{Special Astrophysical Observatory, Russian Academy of Sciences,
Nizhny Arkhyz, Karachai-Circessia 357147, Russia \\ Electronic mail:
akn@sao.ru}
\and
\author{Ulrich Hopp{\footnote[4]{Visiting astronomer,
  Calar Alto German-Spain Observatory.}}}
\affil{Universit\"{a}ts-Sternwarte, Scheiner Str. 1, D-81679, Munich,
Germany\\ Electronic mail:  hopp@usm.uni-muenchen.de}

\begin{abstract}

We present Multiple Mirror Telescope (MMT) and Keck II 
telescope spectrophotometry and 3.5m Calar
Alto telescope $R$, $I$ photometry of the western component of 
the extremely low-metallicity blue compact galaxy SBS 0335--052. 
The components, separated by 22 kpc, appear to be members of  
a unique, physically connected system. It is shown that SBS 0335--052W consists
of at least three stellar clusters and has the same redshift
as SBS 0335--052. The oxygen
abundance in its two brightest knots is  
extremely low, 12 + log (O/H) = 7.22$\pm$0.03 and 7.13$\pm$0.08,
respectively. These values are lower than in SBS 0335--052 and
are nearly the same as those in I Zw 18. 
The $(R-I)$ color profiles are very blue in both galaxies 
due to the combined effects of ionized gas and a young stellar
population emission. We argue that SBS 0335--052W is likely to be 
a nearby, young dwarf galaxy. 

\end{abstract}


\section {Introduction}

    Since its discovery as an extremely low-metallicity galaxy, the
blue compact galaxy (BCG) SBS 0335--052 (SBS -- Second Byurakan
Survey) has been seen as a good candidate to be a nearby, young
dwarf galaxy (Izotov et al. 1990).  This chemically-unevolved galaxy
with oxygen abundance 1/40 the solar value (Terlevich et al. 1992;
Melnick, Heydary-Malayeri \& Leisy 1992; Izotov et al.  1997) is the
second most metal-deficient BCG known, after I Zw 18, with an oxygen
abundance of 1/50 (O/H)$_\odot$ (Skillman \& Kennicutt 1993,
hereafter SK93).  

New evidence in favor of the evolutionary youth of
SBS 0335--052 has been found in subsequent studies. {\sl Hubble Space
Telescope} ({\sl HST}) $V, I$ imaging of this galaxy (Thuan, Izotov \&
Lipovetsky 1997, hereafter TIL97) revealed blue $(V-I)$ colors not
only in the region of current star formation, but also in the
4 kpc diameter extended, low-intensity envelope.  The emission from 
the underlying component, with $(V-I)$ = 0.0--0.2, would seem to arise 
from the combination of ionized gas emission and emission from young 
($\leq$10$^8$ yr) stars (Izotov et al. 1997).  

Another piece of evidence favoring of a young age
for SBS 0335--052 has been presented by Thuan \& Izotov (1997) from {\sl
HST} $UV$ spectrophotometry.  They have shown that this galaxy is a
damped Ly$\alpha$ system with extremely high neutral hydrogen column
density $N$(H I) = 7$\times$10$^{21}$cm$^{-2}$ suggesting a large
amount of unprocessed neutral gas around the galaxy. The heavy element
abundances in neutral gas are several orders of magnitude lower than
those in the ionized gas (Thuan \& Izotov 1997).  

The Very Large Array
(VLA) map of SBS~0335--052 reveals the presence at the same redshift 
of a large extended H I cloud, 64 kpc in size, with a mass two
orders of magnitude larger than that of the observed stars (Pustilnik
et al. 1998).  Two prominent, slightly-resolved H I peaks have been
detected, separated by 22 kpc.  The eastern peak coincides
approximately with SBS 0335--052.  Pustilnik et al. (1997) sought to
identify the western H I peak using the Digital Sky Survey
(DSS) and found an optical counterpart designated as
SBS~0335--052W. The redshift of this faint, slightly elongated compact
galaxy, derived from a 6m telescope optical spectrum, is close to
that of SBS~0335--052, suggesting that these two galaxies and 
the H I cloud form a common system. 

The [N II] and [S II] lines in SBS
0335--052W
were too weak to be detected by Pustilnik et al. (1997) who
concluded that this object may be a low-metallicity young
galaxy.  However, this conclusion is based on a low S/N spectrum in
the $\lambda$$\lambda$4800--7200\AA\ spectral range, so that the
oxygen abundance was not derived. In this paper we present new MMT and
Keck II spectrophotometric observations and 3.5m Calar Alto
telescope CCD $R, I$ photometry, derive for the first time the element
abundances and obtain $R, I$ and $(R-I)$ profiles for SBS 0335--052W.

\section {OBSERVATIONS}
\subsection {Optical spectroscopy}
The MMT optical spectra of SBS 0335--052W were obtained on 1996
January 19 with the Red Channel of the MMT Spectrograph.  We used a
1\arcsec$\times$180\arcsec\ slit; the 300 g mm$^{-1}$ grating provides
a dispersion of 3 \AA\ pixel$^{-1}$ and spectral resolution of
about 10 \AA\ in first order. To avoid second-order
contamination, an L-38 blocking filter was used. The total spectral
range was $\lambda$$\lambda$3700--7300 \AA. The spectra were rebinned
by a factor of 2 along the spatial axis, hence, the spatial
sampling was 0\farcs6 pixel$^{-1}$.  The linear scale is 1\arcsec\ =
263 pc at the distance to SBS 0335--052W of 54.1 Mpc{\footnote[5]{Hubble 
constant $H_0$=75 km s$^{-1}$ Mpc$^{-1}$ is used throughout the text.}}.  The 
total exposure time was 60 min, broken to three exposures of 20 min.  The
slit was oriented north-south (PA=0\arcdeg).  The seeing during the
observations was
$\sim$0\farcs8 and the galaxy was observed at the airmass of 1.25.
The standard star PG 0205+134 was observed for
absolute flux calibration. Spectra of He--Ne--Ar comparison lamps were
obtained before and after each observation to provide wavelength
calibration.

The Keck II telescope optical spectra of SBS 0335--052W were obtained
on 1997 November 10 with the Low Resolution Imaging
Spectrometer (LRIS; Oke et al. 1995), using the 600 g mm$^{-1}$ grating
which provides a
dispersion 1.28 \AA\ pixel$^{-1}$ and a spectral resolution of about 4
\AA\ in first order. The total spectral range covered was
$\lambda$$\lambda$4100--6800 \AA. The 1\arcsec$\times$180\arcsec\ slit was
aligned along the SBS
0335--052W major axis with a position angle P.A. =
--80\arcdeg. No binning along the spatial axis has been done, yielding
a spatial sampling of 0\farcs2 pixel$^{-1}$. The total exposure time
was 90 min, broken to three exposures of 30 min. The seeing during the
observations was around 0\farcs6 and the galaxy was observed at the
airmass of 1.25.  The standard star GD 50 was observed to obtain
an absolute flux calibration. A Ne--Ar comparison spectrum
lamp was obtained after the observation to facilitate wavelength
calibration.
   
The data reduction was carried out at the NOAO headquarters in Tucson
using the IRAF\footnote[6]{IRAF: the Image Reduction and Analysis
Facility is distributed by the National Optical Astronomy
Observatories, which is operated by the Association of Universities
for Research in Astronomy, In. (AURA) under cooperative argeement with
the National Science Foundation (NSF).} software package. Procedures
included bias subtraction, cosmic-ray removal and flat-field
correction using exposures of a quartz incandescent lamp. After
wavelength mapping and night sky background subtraction each frame was
corrected for atmospheric extinction and flux calibrated. The
one-dimensional spectra were extracted from each flux-calibrated frame
using the APALL routine. The extracted spectra were then co-added, not
combined. The cosmic rays hits have been removed manually.  To derive
the sensitivity curves, we have fitted the observed spectral energy
distributions of the standard stars with a high-order polynomial.  
The r.m.s. of the sensitivity response curve around the fit 
was $\sim$ 2\% and $\sim$ 1\% (in linear scale) during MMT and Keck II 
telescope observations respectively.

The MMT spectrum of the brightest 1\arcsec$\times$2\arcsec\ region
of SBS 0335--052W is shown in Figure 1. The two-dimensional
spectrum obtained with LRIS along the major
axis of SBS 0335--052W shows the presence of three emitting regions
with the brightest region at the NW side of the galaxy and two fainter
in the SE regions. One-dimensional spectra of each region in
apertures 1\arcsec$\times$2\arcsec\ each are shown in Figure 2.  In
Figure 3 we show the spatial distribution of the observed flux and
equivalent width of H$\alpha$ emission line. The location of regions
with spectra presented in Figure 2 is shown by vertical lines.  While
in the MMT spectrum the [O III] $\lambda$4363 emission line is barely
detected and is blended with the strong H$\gamma$ emission line,
it is evidently present in the LRIS spectra of two regions which
allows the reliable determination of the electron temperature
and element abundances.

In Table 1, the observed line intensities and intensities corrected
for interstellar extinction for SBS 0335--052W are shown together with
the extinction coefficient $C$(H$\beta$), observed flux of the
H$\beta$ emission line, the equivalent width $EW$(H$\beta$) and
absorption equivalent width $EW$(abs) of Balmer hydrogen lines. We
show 1$\sigma$ upper intensity limits for the lines which have not
been detected.  To correct for extinction we used the Galactic reddening law
of Whitford (1958). The line intensity errors 
listed in Table 1 take into account the noise statistics in the
continuum which implicitly includes the uncertainties of the
data reduction ( flat-field correction, sky subtraction), 
the errors in fitting of the sensitivity response curves and the errors
in placing the continuum and fitting the line profiles with
gaussians. These errors have been propagated to calculate element
abundances. 
The line intensities in the brightest knot derived from
the MMT and Keck observations are in general agreement
implying the correctness of the measurements despite the
difference in slit PA between them.

To derive the element abundances in SBS 0335--052W we adopted a
two-zone photoionized H II region model (Stasi\'nska 1990). The
electron temperature, $T_e$(O III), is derived from the [O
III]$\lambda$4363/($\lambda$4959 + 5007) line intensity ratio using
a five-level atom model, and the electron temperature $T_e$(O
II) from the relation between $T_e$(O II) and $T_e$(O III) fitted by
Izotov, Thuan \& Lipovetsky (1994, hereafter ITL94) for photoionized H
II models by Stasi\'nska (1990). The electron temperature for 
the S III ion is obtained using prescriptions by Garnett (1992).  The
electron number density $N_e$ is derived from the [S II]$\lambda$6717,
6731 lines.  The values of $T_e$(O III), $T_e$(O II), $T_e$(S III) and
$N_e$(S II) are shown in Table 2.  The ionic and total abundances are
derived from Keck II telescope spectra as described by ITL94 and by
Izotov, Thuan \& Lipovetsky (1997, hereafter ITL97) and are shown in
Table 2. However, due to the fact that these spectra cover a 
smaller spectral range, we use the intensities of the [O II]
$\lambda$3727, [Ne III] $\lambda$3868 and [S II]$\lambda$6717, 6731
emission lines in the MMT spectrum, scaled to the LRIS
spectra, to derive oxygen, neon and sulfur abundances.  The oxygen
abundance in both of the brightest regions of SBS 0335--052W is
0.1--0.2 dex smaller than in SBS 0335--052 (Izotov et al. 1997) and is
the same as that in I Zw 18 (SK93). The helium abundance derived from
He I $\lambda$4471 and $\lambda$5876 emission line intensities is in
agreement with He abundances in brighter blue compact galaxies (ITL94,
ITL97). The abundances of other elements are close to the mean values for
extremely metal-deficient galaxies (Thuan, Izotov \& Lipovetsky 1995;
Izotov et al. 1997; Izotov \& Thuan 1998a).
 
\subsection {CCD photometry}

The $R$,$I$ photometric observations of SBS~0335--052W were obtained
during the study of SBS~0335--052 with the prime focus CCD camera on the
3.5m telescope
in Calar Alto on 1991 October 6.  At f/3.5, with the GEC CCD
1152$\times$770 pixels, (22.5 \micron\ pixel$^{-1}$), the field of
view is 7\farcm3$\times$4\farcm9.  Two exposures of 600 s and 900 s
were taken in $R$ and $I$, respectively. Two clusters, M92 and NGC
7790 were observed as photometric standards. The photometric
calibration was made using the data of Christian et al. (1985).  The
observations were done under photometric conditions with a seeing of
FWHM = 1\farcs44 $\pm$ 0\farcs11 and 1\farcs22 $\pm$ 0\farcs02 in $R$
and $I$ respectively.  Airmasses were 1.45 for $R$ and 1.46 for $I$.

All data reduction was done with MIDAS{\footnote[7]{MIDAS is an
acronym for the European Southern Observatory package -- Munich Image
Data Analysis System.}}. The frames were corrected for bias, dark,
flat-field and, additionally, in the $I$ passband, for fringe-pattern.
Unfortunately, the fringe-pattern correction was not very good and
there was some residual structure in both the horizontal and vertical
directions as result of additional noise in the CCD readout. This
structure was corrected by applying median filters separately in the X-
and Y-directions.

Aperture photometry was performed on the standard star observations
using the MAGNITUDE/CIRCLE task with the same aperture for all
stars. The standard star measurements were used for the determination
of the magnitude zero points and the transformation of the
instrumental CCD magnitudes into the Johnson-Kron-Cousins $RI$ system.
For the standard stars, the comparison of the CCD and the catalogued
magnitudes shows a scatter of $\sigma$ = 0.08 mag for $R$ and $\sigma$
= 0.12 mag for $I$. No color term has been included in the
transformation as tests revealed no significant color term was appropriate.

For the photometry of extended objects, we used the package
SURFPHOT as well as dedicated software for adaptive filtering
developed at the Astrophysical Institute of Potsdam (Lorenz et
al. 1993).  The adaptive filter allows one to reduce pixel noise by
3--4 times without loss of spatial resolution of bright cores of stars
and galaxies and to retain the object's total flux as well. The
smoothing scale of the adaptive filter was 11$\times$11 pixels.
Before applying an adaptive filter a special mask was built, where all
bright stars and galaxies were masked out. Such a mask is necessary
for proper determination of noise statistics used by the filter.  The
code used from the package for adaptive
filtration for fitting of the sky background constructs the background
within the
masked regions, avoiding any polynomial approximation.  Instead the
algorithm of ``stretched skin'' is used, which iteratively fills the
background inside the mask by interpolation of that from areas
outside of the mask.  Total magnitudes are $R$ = 19.03 $\pm$
0.09 and $I$ = 19.08 $\pm$ 0.14 for SBS 0335--052W and $R$ =
16.57 $\pm$ 0.08 and $I$ = 16.92 $\pm$ 0.12 for SBS
0335--052. The total magnitude $I$ for SBS 0335--052 is in a good
agreement with total $I$ = 16.88 from TIL97.

Elliptical fitting was performed with FIT/ELL3 in SURFPHOT package
where isophotes of galaxies were fitted by ellipses and the deviations
were analysed by means of Fourier techniques
as is described by Bender \& Moellenhoff (1987).  To construct a
surface brightness profile we used the effective radius as the
geometrical average, $R_{eq}$=$\sqrt{ab}$.  The brightness profile was
decomposed into two components: a gaussian distribution in
the central part and an exponential disk.  For this, the
FIT package was used. Weights in the fitting procedure were
adopted as proportional to 1/$\sigma$, where $\sigma$ is the
instrumental accuracy of the surface brightness profile.

The $R$ images, isophotes and profiles with their decomposition for
SBS 0335--052 and SBS 0335--052W are shown in Fig.4.  The error bars
for surface brightness profiles in both $R$ and $I$ bands are less
than the symbol sizes. Therefore they are shown only on the
$(R-I)$ profiles.  The error bar for the first point shows the
accuracy of the transformation to the standard system.  The error bars
for the remaining points show the instrumental accuracy of the
color and take into account the detector gain and photon statistics
for each frame.  The best linear fits to the outer parts of the $R$
surface brightness distributions are given by
\begin{equation}
   \mu_R = (21.25 \pm 0.05) + (2.03 \pm 0.02)r
\end{equation}
for SBS 0335--052 and
\begin{equation}
    \mu_R = (22.22 \pm 0.12) + (1.60 \pm 0.07)r
\end{equation}
for SBS 0335--052W.

The corresponding exponential scale lengths in the $R$ band 
are $\alpha_R$ = 535$\pm$5 pc and $\alpha_R$ = 422$\pm$18 pc for 
SBS 0335--052 and SBS 0335--052W respectively.

\section {DISCUSSION}

     The discovery of the new dwarf blue compact galaxy SBS
0335--052W poses several important questions: a) How is this
galaxy related to the brighter, extremely low-metallicity galaxy
SBS 0335--052? b) If this system of galaxies is physically connected,
what is its evolutionary state?  Is it an evolved system of
galaxies, or are we observing a system of nearby young dwarf galaxies
during their formation?  c) If this system of galaxies is young, what
kind of constraints should this impose for searches of primeval
galaxies?

     In Table 3 we show the general characteristics of SBS 0335--052
and SBS 0335--052W. Our determination of the heliocentric radial
velocity for SBS 0335--052W is in excellent agreement with the
velocity of SBS 0335--052, and confirms the conclusion of Pustilnik et
al. (1997) that SBS 0335--052 and SBS 0335--052W are physically
related and are two sites of star formation in the large cloud of
neutral gas detected by Pustilnik et al. (1998) in their VLA
observations. Therefore, we suggest that this system is
probably at the very beginning of its evolution.

     More evidence for a common evolution of the two H II
regions in the SBS 0335--052 system comes from spectroscopic
observations of SBS 0335--052W. The spectra of this H II region, shown
in Figures 1 and 2, resemble a spectrum of more moderate excitation as
compared with the spectrum of the brighter companion and shows
stronger low excitation [O II]$\lambda$3727 emission.  

The age of ionizing stellar clusters can be estimated from the equivalent
width of the H$\beta$ emission line. This age estimate may be subject to
uncertainties introduced by the differences in spatial distribution of
ionized gas and ionizing stellar clusters. Mindful of this possible 
source of uncertainty, the lower
equivalent width of H$\beta$ in SBS 0335--052W by factor of $\sim$ 2
implies that its youngest ionizing stellar clusters are older by a
few times 10$^6$ yr than those in SBS 0335--052. However, {\sl HST}
WFPC2 observations of SBS 0335--052 (TIL97, Papaderos et al. 1998) 
have shown the presence of
several super-star clusters spanning in age the range from a few 
to several tens of Myr. The age of the oldest extended
low-intensity stellar population in SBS 0335--052 is $\sim$ 100 Myr
(Izotov et al. 1997; TIL97).  These observations imply that the
assumption of an instantaneous burst of star formation is not
valid. Instead, star formation in SBS 0335--052 during last $\sim$ 100
Myr would be consistent with propagating star formation from NW to SE 
at a velocity $\sim$ 18 km s$^{-1}$ in a region of $\sim$ 1
kpc in size (Papaderos et al. 1998). 
A similar situation is observed in the other known
very metal-deficient galaxy, I Zw 18, where star formation 
likely propagated in the SE direction from the older C component to the
youngest SE component (Izotov \& Thuan 1998b).  

While the
youngest ionizing cluster in the SBS 0335--052W seems to be older than 
that in the SBS 0335--052, the difference in
oxygen abundance suggests that in general SBS 0335-052W might be
younger than its brighter companion.  The oxygen abundance in the two
regions of SBS 0335--052W, 12 + log (O/H) = 7.22$\pm$0.03 and
7.13$\pm$0.08, is the lowest known for dwarf emission-line galaxies,
and it is comparable with 12 + log (O/H) = 7.17 -- 7.26 derived for I
Zw 18 (SK93).

      In Figure 4 we show the $R, I$ and $(R-I)$ profiles for both SBS
0335--052 and SBS 0335--052W obtained with the 3.5m telescope. The
surface brightness distribution in $I$ for SBS 0335--052 is in a good
agreement with that obtained by TIL97 and the total $I$ magnitude
derived from both observations is coincident within 0.05 mag. The $R$
and $I$ profiles for SBS 0335--052 and SBS 0335--052W at large
distances are fitted well by an exponential.  The $(R-I)$ color in
both HII regions gets redder with increasing radius.  Izotov et
al. (1997) have shown that the $(V-I)$ color of the extended
stellar emission is strongly modified by the emission of ionized gas.
It is evident from Fig. 3b that the $(R-I)$ color in SBS 0335--052W is
affected by the presence of ionized gas emission but to a lesser
extent than in SBS 0335--052 where gaseous emission is
stronger. The $R$ passband is more subject to the influence of
gaseous emission due to the strong H$\alpha$ emission which
reaches its maximum in the region 3\farcs6 SE of the
brightest part of the galaxy. In this region $R$ is
$\sim$ 0.5 mag brighter due to the H$\alpha$ emission line
contribution, while in the brightest region the brightening due to the
presence of the H$\alpha$ emission is smaller, $\sim$ 0.3
mag. However, the equivalent width of H$\alpha$ is small at
distances $r$$>$3\arcsec\ NW of the center of SBS~0335--052W 
and at distances $r$$>$5\arcsec\ SE of the center
(Fig. 3b). Therefore, corrections for gaseous emission
in the outer part of SBS 0335--052W are small, not exceeding $\sim$
0.17 mag, and being even less than 0.1 mag in regions with
$EW$(H$\alpha$) $\leq$ 200\AA.

To derive the age of the extended stellar component in the SBS 0335--052W,
we have
calculated a grid of spectral energy distributions (SED) for stellar 
populations with ages between 10 Myr and 20 Gyr using isochrones of stellar
parameters from Bertelli et al. (1994) for metallicities 1/20 of the solar
value
and the compilation of stellar atmosphere models from Lejeune, Cuisiner \&
Buser (1998). We adopt an initial mass function (IMF) with 
a slope --2.35, and lower and upper mass limits of 
0.6$M_\odot$ and 100$M_\odot$, respectively. 

The resulting calculated broad-band colors
as a function of age of the stellar population produced during a single
burst are shown in Table 4. The sharp change of $(R-I)$ between 
0.17 at log $t$=8.0 and 0.36 at log $t$=8.1 is caused by the appearance of
the first asymptotic giant branch stars, where $t$ is
expressed in yr. The $(V-I)$ and $(V-K)$ colors are even more
sensitive to the appearance of the first asymptotic giant branch stars
and, therefore, they are better discriminators between young and old
stellar populations.  

It follows from Table 4 that the observed 
$(R-I)$ = 0.2 in the outer envelope of SBS 0335--052W can be
explained only by the presence of a stellar population with age
$\leq$100 Myr.  The assumption of a continuous star formation
rate in the galaxy with an age $\leq$ 100 Myr does not change its colors
appreciably due to the weak dependence of colors on age at $t$ = 10 --
100 Myr (Table 4) which is the most likely age for the extended stellar
component in SBS 0335--052 and its western counterpart.  Of course, we
cannot exclude completely the presence of an old stellar
population simply because the young stellar population dominates the
radiation. However, if present, the extended old population in the SBS
0335--052 is less significant than in the majority of BCGs
where it is easily detected.  Hence, the $R, I$
photometric results contradict neither the hypothesis of a young age for
SBS~0335--052 and SBS~0335--052W, nor their
common origin during the last $\leq$100 Myr. However, as the
$(R-I)$ color is a weak discriminator of the age of a
stellar population, photometric observations in other broad bands
are necessary to confirm this hypothesis.

The existence of two young regions of star formation separated by 24
kpc poses the problem of identifying a mechanism for synchronizing star
formation. The velocity distribution of neutral gas in SBS 0335--052
shows no jumps or discontinuities and resembles the velocity
distribution in a rotating gaseous disk (Pustilnik et al. 1998). No
stellar or ionized gaseous optical emission is detected
between both the components (Papaderos et al. 1998).  These observations
argue against models where star formation in two widely separated
regions is caused by the collision of two gaseous clouds or by tidal
effects. 

One plausible mechanism for synchronizing 
star formation in two widely separated regions could be 
the contraction of a protogalaxy. Adopting the number density of 
the neutral gas $\sim$0.1--1 cm$^{-3}$ from TIL97, we derive a
free-fall time of $\leq$100 Myr, in agreement with our estimates of the
age of the stellar population.

      The SBS 0335--052 system has several similarities
with the dwarf galaxy associated with the large H I cloud in
Virgo (1225+01) discovered by Giovanelli \& Haynes
(1989).  However, the SBS 0335--052 system is characterized by one
order of magnitude larger luminosity and star formation rate.  The H I
distribution of 1225+01 shows two peaks separated by 15 arcmin, or 45
kpc, if a distance of 10 Mpc is adopted. The optical knots resembling
H II regions coincide with one of the peaks; the other peak, however, has
no optical counterpart. Salzer et al. (1991) presented a detailed
optical imaging and spectroscopic study of the dwarf irregular galaxy
located at the main peak of the H I cloud. They found that the entire
galaxy is very blue. Nebular abundances derived from the spectroscopic
data reveal that this object is relatively unevolved chemically,
though it has an oxygen abundance 12 + log (O/H) = 7.66, two to three
times larger than in the SBS 0335--052 system. Furthermore,
Salzer et al.  found that only a tiny fraction (0.02\% -- 0.60\%) of
the mass in the NE H I clump of 1225+01 is in the form of stars. A similar
value is found in the SBS 0335--052 system (Izotov et al. 1997).
Based on the models of chemical and color evolution, Salzer et
al.  concluded that the galaxy associated with 1225+01 H I cloud is
still undergoing formation and the stellar population of this
galaxy is likely no older than roughly 1 Gyr.

   If our conclusion about the youth of SBS 0335--052 is correct, we
can compare properties of this young galaxy with theoretical
predictions and properties of high-redshift galaxies which are often
considered primeval.  In SBS 0335--052, the Ly$\alpha$
emission line is not observed (TIL97, Thuan \& Izotov 1997), contrary
to theoretical predictions for young galaxies (Charlot \& Fall
1993). The small mass, $\sim$ 10$^7$$M_\odot$, of the older stellar
population in SBS 0335--052 suggests that the star formation rate
during the first 100 Myr is only 0.1--1 $M_\odot$ yr$^{-1}$. During
this period only 1\% of gaseous mass is transformed to stars, contrary
to predictions of several models of galaxy formation (Meier 1976;
Baron \& White 1987; Lin \& Murray 1992), where a significant part of
the gas converts to stars during the collapse time. Probably, the
formation of galaxies, or at least dwarf galaxies, is a more quiescent
process with moderate conversion of gas into stars. We expect that a
detailed study of the SBS 0335--052 system could provide constraints on
models of galaxy formation and could provide useful guidelines with 
which to search for primeval galaxies at large redshifts.

\acknowledgements
 V.A.L., Y.I.I. and A.Y.K. thank R. Green and the staff of NOAO for 
their kind hospitality. The authors are very grateful to Simon Pustilnik for
stimulating discussions. Partial financial support for this international 
collaboration was made possible by INTAS collaborative research grant
No 94--2285. U.H. acknowledges the support by the SFB 375 of
the Deutsche Forschungsgemeinschaft.
C.B.F. and F.H.C. acknowledge support from NSF grant AST 93-20715, for which
they are very grateful.


\clearpage

\figcaption[fig1.ps]{The observed spectrum of SBS 0335--052W
obtained with Multiple Mirror Telescope in an 1\arcsec\ $\times$ 2\arcsec\ 
aperture.}

\figcaption[fig2.ps]{The Keck II telescope spectra of three regions in
SBS 0335--052W aligned in the direction from the SE (a) to the NW
(c). All spectra have been extracted in the 1\arcsec\ $\times$
2\arcsec\ apertures.}

\figcaption[fig3.ps]{The spatial distribution of the observed flux (a)
and equivalent width (b) of the H$\alpha$ emission line
in 1\arcsec\ $\times$ 0\farcs6 apertures. The location of
three regions with spectra displayed in Figure 2 are marked by vertical
lines in Figure 3a. The linear scale is 1\arcsec\ = 263 pc.}

\figcaption[fig4.ps]{ a) $R$-band images of SBS 0335--052 and SBS
0335--052W, obtained with
the 3.5m Calar Alto telescope. North is up, East is left (scales
as below);
b) Isophotes in the $R$ passband for SBS 0335--052 and SBS 0335--052W.
The outermost isophote is 26 mag arcsec$^{-2}$,
the step is 0.5 mag arcsec$^{-2}$, the linear scale is 1\arcsec\ = 263 pc;
c) The $R, I$ and $(R-I)$ profiles of SBS 0335--052 and SBS 0335--052W.
The decomposition of $R$ profiles on central starburst component and 
extended low surface brightness exponential profiles is shown by dotted
lines. The error bars are shown only for the $(R-I)$ profiles.
The error bar for the first point shows the accuracy of the transformation to
the standard system.
The error bars for the rest points show the instrumental accuracy.
The brighter galaxy is bluer in the central region, but the colors
for both galaxies are the same in outer regions.} 

\clearpage
%
%
\begin{deluxetable}{lcccccccc}
\small
\tablenum{1}
\tablecolumns{9}
\tablewidth{0pt}
\tablecaption{Emission Line Intensities}
\tablehead{
\colhead{}& \multicolumn{5}{c}{Keck II}&\colhead{}&\multicolumn{2}{c}{MMT} 
\\ \cline{2-6} \cline{8-9} \\
\colhead{Ion}& \multicolumn{2}{c}{Western knot}&\colhead{}&
\multicolumn{2}{c}{Eastern knot}&\colhead{}&\multicolumn{2}{c}{Western knot} \\ \cline{2-3} \cline{5-6} \cline{8-9} \\
\colhead{}&\colhead{$F$($\lambda$)/$F$(H$\beta$)}
&\colhead{$I$($\lambda$)/$I$(H$\beta$)}&\colhead{}
&\colhead{$F$($\lambda$)/$F$(H$\beta$)}
&\colhead{$I$($\lambda$)/$I$(H$\beta$)}&\colhead{}
&\colhead{$F$($\lambda$)/$F$(H$\beta$)}
&\colhead{$I$($\lambda$)/$I$(H$\beta$)} }
\startdata
 3727\ [O II]        &\nodata        & \nodata        &&\nodata        &\nodata        &&0.652$\pm$0.027&0.698$\pm$0.031 \nl
 3868\ [Ne III]      &\nodata        & \nodata        &&\nodata        &\nodata        &&0.146$\pm$0.014&0.154$\pm$0.015 \nl
 3889\ He I + H8     &\nodata        & \nodata        &&\nodata        &\nodata        &&0.123$\pm$0.012&0.182$\pm$0.023 \nl
 3968\ [Ne III] + H7 &\nodata        & \nodata        &&\nodata        &\nodata        &&0.118$\pm$0.015&0.178$\pm$0.027 \nl
 4101\ H$\delta$     &0.177$\pm$0.005& 0.254$\pm$0.008&&0.200$\pm$0.012&0.219$\pm$0.016&&0.197$\pm$0.013&0.250$\pm$0.020 \nl
 4340\ H$\gamma$     &0.405$\pm$0.009& 0.458$\pm$0.011&&0.434$\pm$0.017&0.446$\pm$0.019&&0.434$\pm$0.019&0.479$\pm$0.024 \nl
 4363\ [O III]       &0.036$\pm$0.002& 0.035$\pm$0.002&&0.044$\pm$0.007&0.043$\pm$0.007&&$<$0.05\tablenotemark{a}        &$<$0.05         \nl
 4471\ He I          &0.037$\pm$0.002& 0.036$\pm$0.002&&0.040$\pm$0.007&0.040$\pm$0.007&&$<$0.03        &$<$0.03         \nl
 4686\ He II         &$<$0.01        & $<$0.01        &&$<$0.02        &$<$0.02        &&$<$0.03        &$<$0.03         \nl
 4861\ H$\beta$      &1.000$\pm$0.020& 1.000$\pm$0.021&&1.000$\pm$0.030&1.000$\pm$0.031&&1.000$\pm$0.037&1.000$\pm$0.038 \nl
 4959\ [O III]       &0.489$\pm$0.010& 0.460$\pm$0.010&&0.474$\pm$0.017&0.467$\pm$0.019&&0.451$\pm$0.019&0.435$\pm$0.019 \nl
 5007\ [O III]       &1.452$\pm$0.028& 1.366$\pm$0.028&&1.379$\pm$0.040&1.358$\pm$0.040&&1.308$\pm$0.047&1.259$\pm$0.047 \nl
 5876\ He I          &0.103$\pm$0.003& 0.094$\pm$0.003&&0.095$\pm$0.008&0.094$\pm$0.008&&0.112$\pm$0.009&0.102$\pm$0.008 \nl
 6300\ [O I]         &0.015$\pm$0.002& 0.013$\pm$0.002&&0.017$\pm$0.004&0.017$\pm$0.004&&$<$0.02        &$<$0.02         \nl
 6312\ [S III]       &0.006$\pm$0.002& 0.005$\pm$0.002&&$<$0.02        &$<$0.02        &&$<$0.02        &$<$0.02         \nl
 6563\ H$\alpha$     &3.069$\pm$0.059& 2.769$\pm$0.061&&2.720$\pm$0.073&2.685$\pm$0.079&&3.129$\pm$0.107&2.759$\pm$0.106 \nl
 6583\ [N II]        &0.025$\pm$0.002& 0.022$\pm$0.002&&0.022$\pm$0.006&0.022$\pm$0.006&&\nodata\tablenotemark{b}        &\nodata         \nl
 6678\ He I          &\nodata        & \nodata        &&\nodata        &\nodata        &&$<$0.03        &$<$0.03         \nl
 6717\ [S II]        &\nodata        & \nodata        &&\nodata        &\nodata        &&0.071$\pm$0.008&0.062$\pm$0.008 \nl
 6731\ [S II]        &\nodata        & \nodata        &&\nodata        &\nodata        &&0.049$\pm$0.006&0.043$\pm$0.005 \nl \nl
 $C$(H$\beta$) dex    &\multicolumn {2}{c}{0.075$\pm$0.025}&\colhead{}&\multicolumn {2}{c}{0.000$\pm$0.035}&\colhead{}&\multicolumn {2}{c}{0.135$\pm$0.044} \nl
 $F$(H$\beta$)\tablenotemark{c} &\multicolumn {2}{c}{1.95$\pm$0.03}&\colhead{}&\multicolumn{2}{c}{0.55$\pm$0.02}&\colhead{}&\multicolumn{2}{c}{1.76$\pm$0.05} \nl
 $EW$(H$\beta$)\ \AA &\multicolumn {2}{c}{99$\pm$1}&\colhead{}&\multicolumn {2}{c}{140$\pm$2}&\colhead{}&\multicolumn {2}{c}{86$\pm$1} \nl
 $EW$(abs)\ \AA      &\multicolumn {2}{c}{5.7$\pm$0.2}&\colhead{}&\multicolumn{2}{c}{ 1.9$\pm$0.7}&\colhead{}&\multicolumn{2}{c}{ 2.4$\pm$0.5} \nl
\enddata
\tablenotetext{a}{blend with H$\gamma$ $\lambda$4340.}
\tablenotetext{b}{blend with H$\alpha$ $\lambda$6563.}
\tablenotetext{c}{in units of 10$^{-15}$ ergs\ s$^{-1}$cm$^{-2}$.}
\end{deluxetable}

\clearpage
%
%
\begin{deluxetable}{lcc}
\small
\tablenum{2}
\tablecolumns{3}
\tablewidth{300pt}
\tablecaption{Ionic and Total Element Abundances\tablenotemark{a}}
\tablehead{
\colhead{Property}&\colhead{Western knot}&\colhead{Eastern knot} }
\startdata
$T_e$(O III)(K)                                      &17,200$\pm$600      &19,300$\pm$1,900     \nl
$T_e$(O II)(K)                                       &14,700$\pm$500      &15,500$\pm$1,500     \nl
$T_e$(S III)(K)                                      &16,000$\pm$500      &17,700$\pm$1,600     \nl
$N_e$(S II)(cm$^{-3}$)                               &    10$\pm$10       &    10$\pm$10        \nl \nl
O$^+$/H$^+$($\times$10$^5$)\tablenotemark{b}         &0.60~~$\pm$~0.06    &0.51~~$\pm$~0.13     \nl
O$^{++}$/H$^+$($\times$10$^5$)                       &1.08~~$\pm$~0.09    &0.85~~$\pm$~0.20     \nl
O/H($\times$10$^5$)                                  &1.68~~$\pm$~0.11    &1.35~~$\pm$~0.23     \nl
12 + log(O/H)                                        &7.22~~$\pm$~0.03    &7.13~~$\pm$~0.07     \nl \nl
N$^+$/H$^+$($\times$10$^7$)                          &1.72~~$\pm$~0.17    &1.52~~$\pm$~0.38     \nl
ICF(N)                                               &2.81\,~~~~~~~~~~    &2.67\,~~~~~~~~~~     \nl 
log(N/O)                                             &--1.54~~$\pm$~0.07~~&--1.53~~$\pm$~0.19~~ \nl \nl
Ne$^{++}$/H$^+$($\times$10$^5$)\tablenotemark{b}     &0.24~~$\pm$~0.03    &\nodata              \nl
ICF(Ne)                                              &1.55\,~~~~~~~~~~    &\nodata              \nl 
log(Ne/O)                                            &--0.65~~$\pm$~0.07~~&\nodata              \nl \nl
S$^+$/H$^+$($\times$10$^7$)\tablenotemark{b}         &1.09~~$\pm$~0.11    &\nodata              \nl
S$^{++}$/H$^+$($\times$10$^7$)                       &2.15~~$\pm$~0.64    &\nodata              \nl
ICF(S)                                               &1.29\,~~~~~~~~~~    &\nodata              \nl 
log(S/O)                                             &--1.60~~$\pm$~0.08~~&\nodata              \nl \nl
$\gamma$(He I $\lambda$4471)\tablenotemark{c}        &0.002~~~~~~~~~      &0.004~~~~~~~~~       \nl
He$^+$/H$^+$($\lambda$4471)                          &0.076$\pm$0.005     &0.086$\pm$0.016      \nl
$\gamma$(He I $\lambda$5876)\tablenotemark{c}        &0.004~~~~~~~~~      &0.005~~~~~~~~~       \nl
He$^+$/H$^+$($\lambda$5876)                          &0.078$\pm$0.002     &0.080$\pm$0.007      \nl
He$^{++}$/H$^+$($\lambda$4686)                       &$<$10$^{-4}$        &$<$10$^{-4}$         \nl
ICF(He)                                              &1.000~~~~~~~~~      &1.000~~~~~~~~~       \nl
He/H (weighted mean)                                 &0.078$\pm$0.002     &0.081$\pm$0.006      \nl
$Y$ (weighted mean)                                  &0.238$\pm$0.007     &0.244$\pm$0.020      \nl
\enddata
\tablenotetext{a}{Abundances are derived based on Keck II telescope spectra.}
\tablenotetext{b}{The abundance is calculated from observed relative intensity of ion line derived from MMT spectrum.}
\tablenotetext{c}{(1+$\gamma$)$^{-1}$ is a correction factor for collisional enhancement of He I emission line.}
\end{deluxetable}

\clearpage

\begin{deluxetable}{lrr}
\tablenum{3}
\tablecolumns{3}
\tablewidth{300pt}
\tablecaption{Observed and Derived Parameters for the SBS 0335--052 system}
\tablehead{
\colhead{Parameter}&\colhead{0335--052W}&\colhead{0335--052} }
\startdata
$\alpha$(2000)                           &03$^h$37$^m$38\fs4
&03$^h$37$^m$44\fs0     \nl
$\delta$(2000)
&--05\arcdeg02\arcmin36\arcsec&--05\arcdeg02\arcmin39\arcsec  \nl
$R$ mag\tablenotemark{a}                 &19.03$\pm$0.09
    &16.57$\pm$0.08  \nl
$R-I$ mag\tablenotemark{a}               & --0.05$\pm$0.15
    & --0.35$\pm$0.14  \nl
$V_{opt}$ km s$^{-1}$                    &4069$\pm$20\tablenotemark{a}
&4060$\pm$12 \nl
$V_{\rm HI}$ km s$^{-1}$\tablenotemark{b}&4017$\pm$~5
&4057$\pm$~5 \nl
$D$ Mpc                                  &54.1
&54.1 \nl
$M_R$ mag                                & --14.64                     &
--17.10 \nl
$R_{26}$ kpc                             & 1.42                        &
2.50 \nl
\enddata
\tablenotetext{a}{This paper.}
\tablenotetext{b}{Pustilnik et al. (1998).}
\end{deluxetable}

\clearpage

\begin{deluxetable}{rccccc}
\tablenum{4}
\tablecolumns{6}
\tablewidth{350pt}
\tablecaption{Synthetic colors of single stellar population with
$Z$=$Z_\odot$/20}
\tablehead{
\colhead{$\log$ $t$\tablenotemark{a}}&\colhead{$(U-B)$}&\colhead{$(B-V)$}
&\colhead{$(V-I)$}&\colhead{$(R-I)$}&\colhead{$(V-K)$} }
\startdata
 7.0&--1.00&--0.15~\, &--0.03~\, &--0.08~\, &--0.31~\, \nl
 7.5&--0.72&  0.01    &  0.35    &  0.13    &  0.73 \nl
 8.0&--0.49&  0.07    &  0.43    &  0.17    &  0.83 \nl
 8.1&--0.41&  0.19    &  0.79    &  0.36    &  1.60 \nl
 8.5&--0.25&  0.23    &  0.83    &  0.38    &  1.66 \nl
10.0&--0.23&  0.65    &  1.14    &  0.48    &  2.06 \nl
\enddata
\tablenotetext{a}{Age $t$ is in yr.}
\end{deluxetable}

\clearpage
\begin{figure*}
\figurenum{1}
\epsscale{2.0}
\plotfiddle{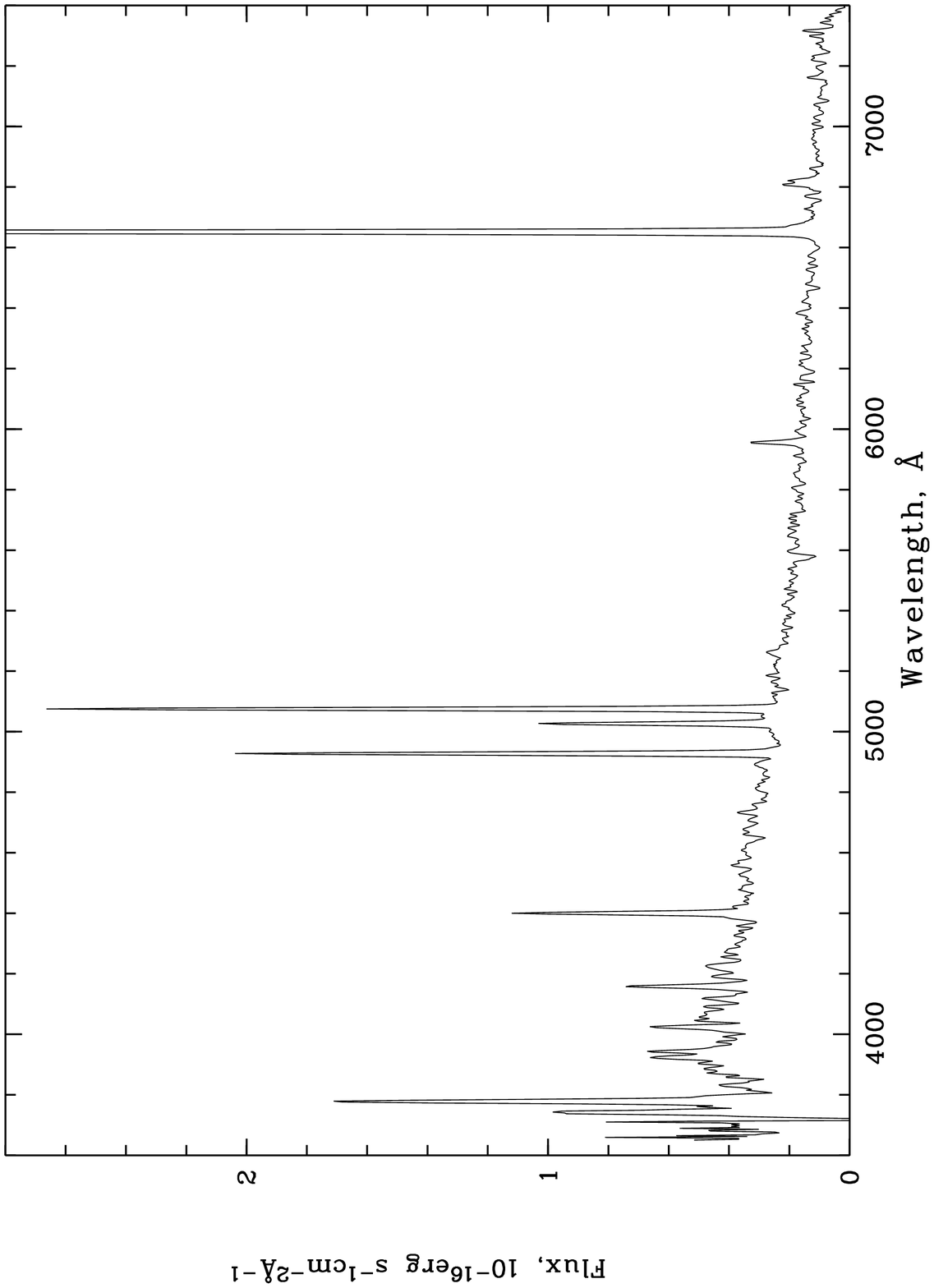}{0.cm}{0.}{85.}{85.}{-270.}{-370.}
\end{figure*}

\clearpage
\begin{figure}
\figurenum{2}
\epsscale{1.6}
\plotfiddle{fig2.eps}{0.cm}{0.}{85.}{85.}{-160.}{-360.}
\end{figure}

\clearpage
\begin{figure*}
\figurenum{3}
\epsscale{1.8}
\plotfiddle{fig3.eps}{0.cm}{0.}{85.}{85.}{-300.}{-360.}
\end{figure*}

\clearpage
\begin{figure*}
\figurenum{3}
\epsscale{1.8}
\plotfiddle{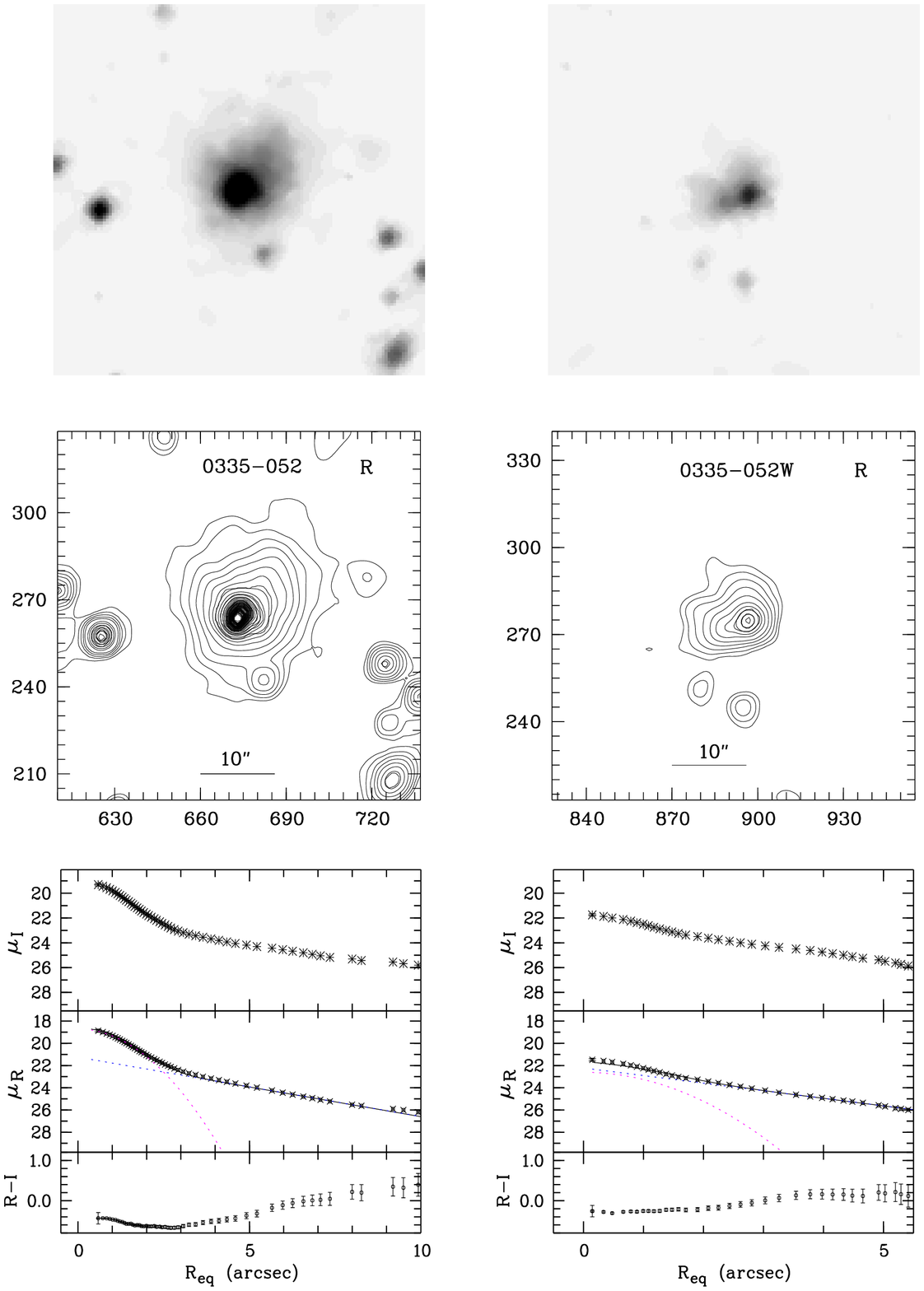}{0.cm}{0.}{85.}{85.}{-280.}{-330.}
\end{figure*}

\end{document}